\documentstyle[11pt,paspconf,psfig,epsf]{article}

\begin{document}

\title{Redshifted 21cm Line Absorption by Intervening Galaxies}
\author{F.H. Briggs}
\affil{Kapteyn Astronomical Institute, P.O. Box 800, 9700 AV
  Groningen, The Netherlands }

\begin{abstract}
The present generation of radio telescopes, combined with
powerful new spectrometers, is opening a new age of redshifted
radio absorption-line studies.  Outfitting of arrays of antennas,
such as the European VLBI Network and the upgraded VLA, 
with flexibly tuned receivers,
will measure sizes and kinematics of intervening galaxies as a
function of cosmic time. 
\end{abstract}


\section{The Past}

Shortly after the discovery that quasars were highly luminous objects
located at cosmological distances, Wagoner (1967), Bahcall \& Spitzer (1969), 
and Gunn \& Peterson (1965) recognized the importance of absorption 
arising in intervening galaxies, extended halos and intergalactic matter,
along with use of the 21cm line as a probe of neutral gas (Bahcall and Ekers 1969).
Quasar absorption lines have subsequently proven to be a gold mine
of information on the evolution of the gaseous content of the Universe.
The following subsection is a reminder that the 21cm line techniques 
can address the epoch during which mass is being most vigorously 
redistributed and galaxies are being assembled.

\subsection{The Distant Past: Comoving Density of Neutral Gas}

A detectable 21cm line optical depth, $\tau_{21}$, 
requires a high column density of neutral gas, 
$$N_{HI}\approx 2{\times}10^{21}\left(\frac{\Delta V}{10\;{\rm km/s}}\right)
     \left(\frac{T_s}{100\;{\rm K}}\right)\tau_{21}\:{\rm cm}^{-2}.$$
For the ranges of spin temperatures $T_s$ and velocity widths $\Delta V$
that are typical of quasar absorption line systems (see Lane these
proceedings), $N_{HI}$ are of order $10^{21}$ cm$^{-2}$ or more.  Gas
clouds of such high column density seen in absorption against quasars
are drawn from the extreme high $N_{HI}$ end of the column density
distribution (cf Petitjean et al 1993)  identified through
absorption in the ultraviolet
Lyman-$\alpha$ line. The high column
densities of $N_{HI}>10^{20}$cm$^{-2}$, which are typical of lines of sight
through the cool disks of nearby late-type galaxies, are named damped
Lyman-$\alpha$ systems (DLa's) after their distinctive absorption profile;
these systems
form the dominant repositories of neutral gas detected by
the absorption-line method (Wolfe et al 1995, Lanzetta et al 1995).
The exact nature of these absorption systems remains to be determined, 
but it is clear that, by tracing   the confining gravitational potentials
that are necessary keep hydrogen neutral in the face of ionizing background
radiation,
they can sense the evolution of galaxies as a function of redshift.

Optical spectroscopy has been the most efficient method for identifying
the DLa's, which make 
interesting targets for 21cm absorption against background radio 
quasars.  This will soon change with the advent of new broadband radio
receiving systems and spectrometers.  The optical spectroscopy is most
efficient for redshifts above $z\approx 1.6$ for which Lyman-$\alpha$\
is shifted to wavelengths that can be observed from the ground.  This excludes
the most recent ${\sim}2/3$ of the age of the universe from study, except
through space based telescopes. The DLa studies must also concentrate on
bright optical quasars, which complicate  the identification of the
intervening absorber and study of its environment, since the bright
quasar continuum can overwhelm the intervenor. In the near future,
identification of 21cm absorbers
in radio spectral surveys will find systems against high redshift radio
galaxies with weak optical continua, thus facilitating the optical
follow up, as well as providing a greater likelihood that the background
radio structures will be extended and allow mapping of the intervenor
in absorption using radio interferometers.

\begin{figure}
\psfig{figure=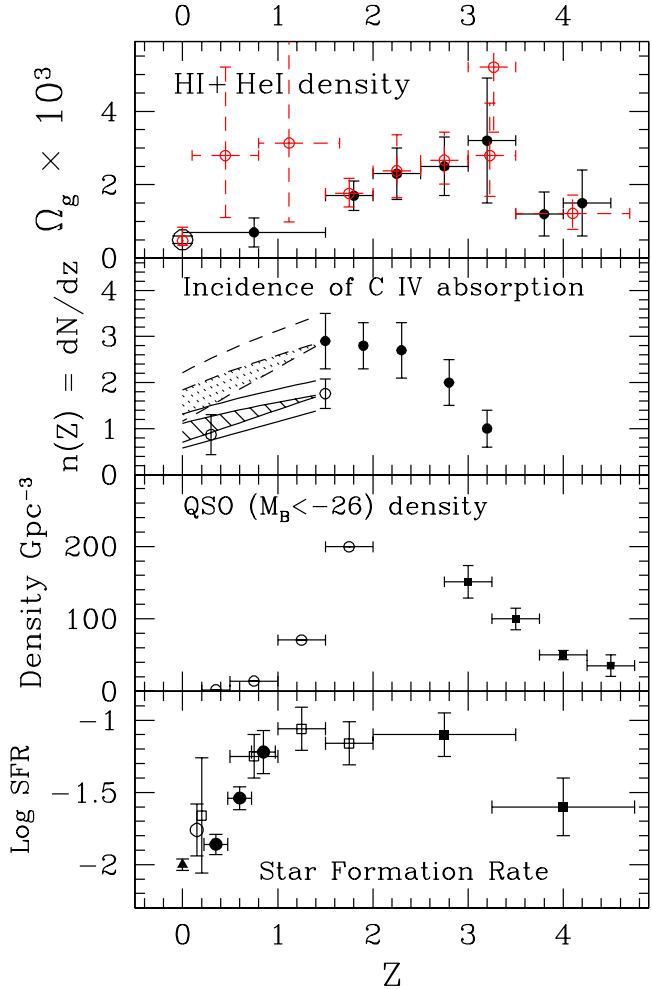,width=14cm,angle=0}
\caption{Cosmological density of neutral gas, incidence 
of CIV absorption, comoving density of luminous QSOs, and mean
star formation rate as a function of redshift. 
{\it Top panel.}  Mean cosmological density of neutral gas,
$\Omega_g$, normalized to the critical density (Storrie-Lombardi et al
1996; Rao et al 1995 ($z=0$); Lanzetta et al 1995, Turnshek 1998)
{\it Upper middle panel.} Number of CIV metal-line absorption systems per unit redshift, $n(z)$ (Steidel 1990); $z=0.3$ point from
Bahcall et al 1993). Filled points
from Steidel indicate rest frame equivalent widths 
$W_{rest}({\lambda}1548) > 0.15$ \AA; open points are for 
$W_{rest}({\lambda}1548) >0.3 $ \AA. 
Hatched areas indicate the range ($0<q_o<1/2$) for unevolving
cross sections since $z=1.5$, beyond which redshift CIV can be
measured with ground-based telescopes. 
{\it Lower middle panel.} Comoving density of optically selected QSOs:
filled squares from Schmidt et al 1994; open circles from Hewitt et al 1993).
$H_o = 50 $km~s$^{-1}$Mpc$^{-1}$, $q_o =1/2$
{\it Bottom panel.} Comoving star formation rate density 
$M_{\odot}$yr$^{-1}$Mpc$^{-3}$ from Madau (1998) and references therein.}
\label{fnct_z.fig}
\end{figure}

The optical surveys for DLa systems have provided convincing evidence
that there was a substantially higher cosmological density of neutral
gas, $\Omega_g$,
 at redshifts 2.5 to 3.5 than there is at the present. The evidence
compiled by Wolfe et al (1995), Lanzetta et al (1995), Storrie-Lombardi
et al (1996), and Rao and Turnshek (1998) is presented
in Fig.~\ref{fnct_z.fig}, where it can be seen that $\Omega_g$ is poorly
constrained for $0.2 <z<1.6$. The radio wavelength instrumentation 
currently under construction
will be especially effective at defining the neutral gas evolution in this
intermediate redshift regime (see \S 3).

A variety of evidence establishes that galaxies undergo substantial
evolution during the
epoch $z\approx 0$ to 5 that is sensed through quasar absorption lines.
Fig.~\ref{fnct_z.fig} shows that (1) $z\approx 3$ is not only
the peak in $\Omega_g(z)$ but also marks the onset of
CIV absorption-lines, which are thought to arise in extended, metal-rich,
ionized halos around galaxies, (2) redshifts around 2.5 mark the peak in
the comoving number density of the most luminous quasars, and (3) the 
star formation rate in distant galaxies rose
from $z\approx$ 5 to $z\approx$ 1 but has since declined by
about a factor of 10 to its current rate.  The decline in star formation
rate is expected correlate with the decline in neutral gas density,
as the reservoir of material available for forming stars is
consumed (cf Lanzetta et al 1995, Fall et al 1996).  Together, these
indicators say that $z\approx 5$ to $z\approx 1$ is a period of vigorous
redistribution of mass,  as gravitational potentials form and confine hydrogen
to drive star formation,
metal-rich halos are polluted with the remains of evolved stars, and  
active nuclei are fueled with maximum effectiveness.  21cm line observations
of this epoch will sense kinematics in evolving galactic potentials during
times when neutral gas masses exceed the luminous stellar mass.
 
\begin{figure}
\hsize 1.03in
\hglue 10cm\epsfxsize=\hsize\epsffile{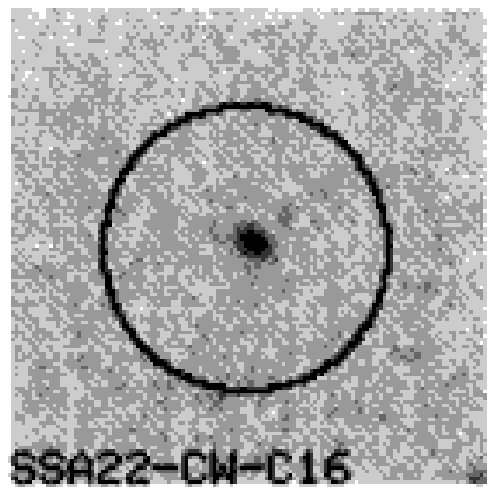}
\hsize 1.5in
\vglue .2in
\hglue 9.4cm\epsfxsize=\hsize\epsffile{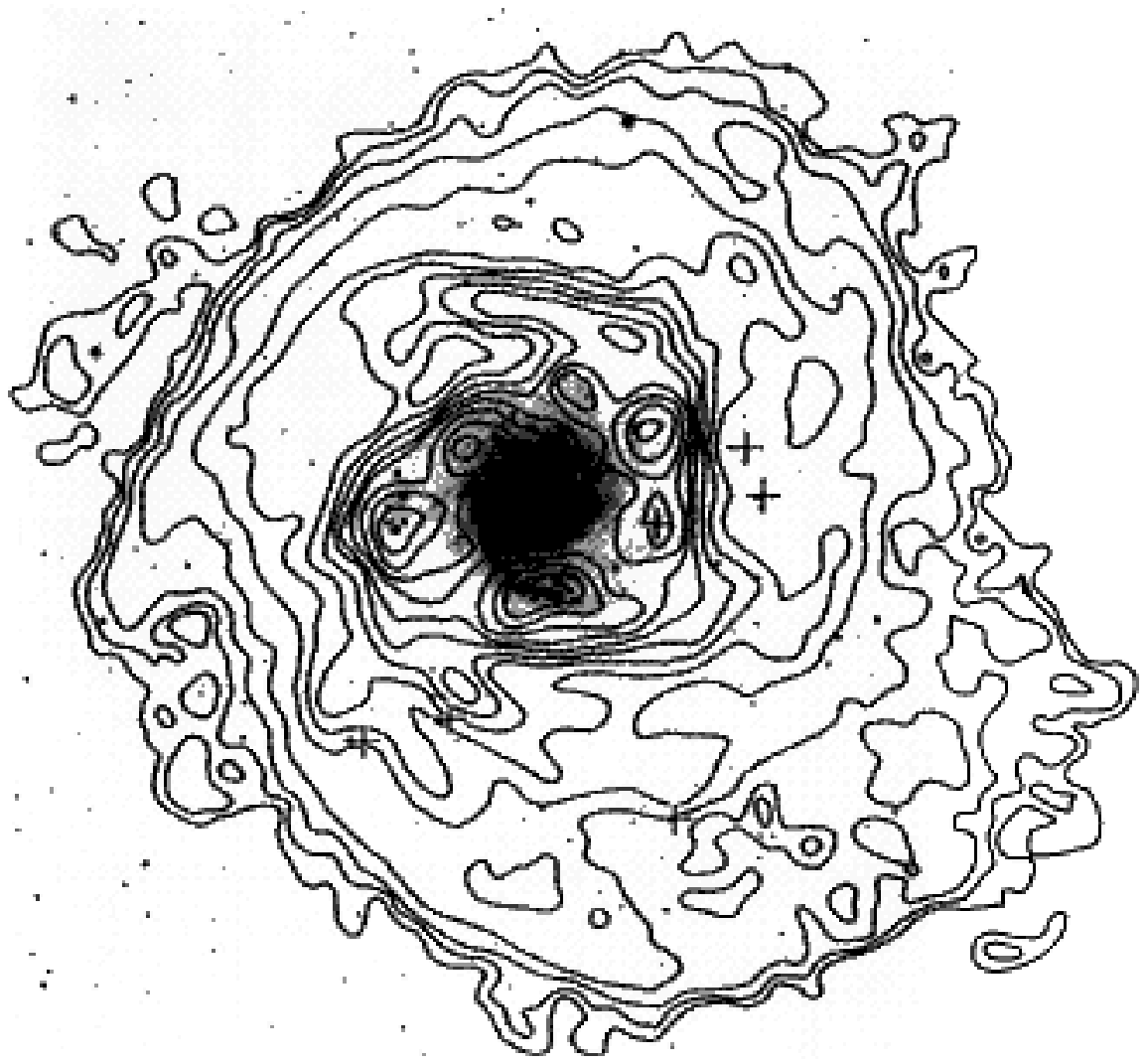}
\hsize 4.5in
\vglue -3.4in
\hglue -.6cm\epsfxsize=\hsize\epsffile{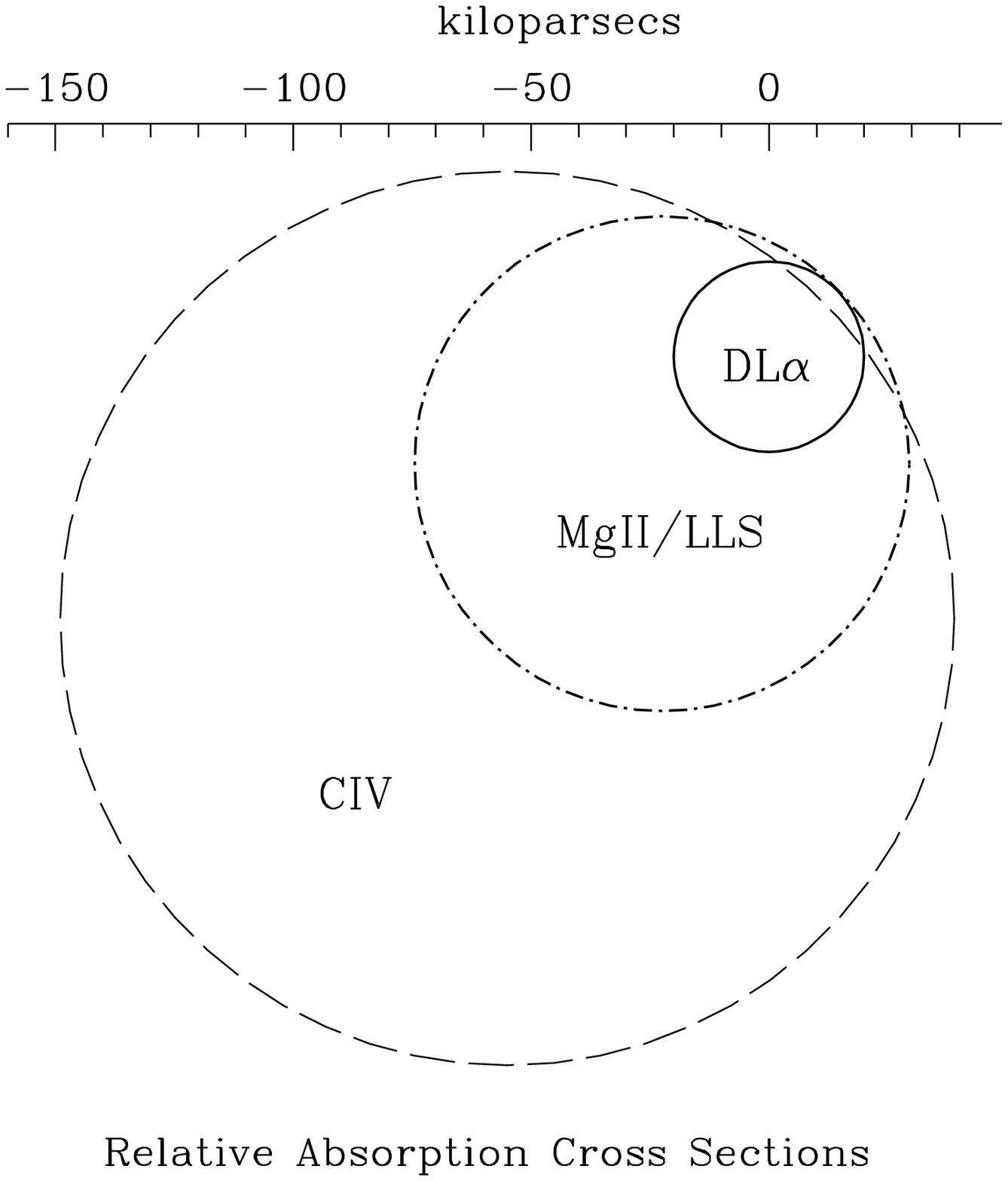}
\hsize 13.5cm
\caption{ Comparison of quasar absorption-line
cross sections for CIV, MgII-Lyman Limit, and damped
Lyman$-\alpha$ lines with the physical size of the optical emission
from  a color-selected galaxy at $z\approx 3$ {\it top right} 
(Giavalisco et al 1996a) and the HI extent of
a nearby, large $L\sim L_*$ galaxy M74=NGC628 {\it lower right} (Kamphuis \&
Briggs 1993). The absorption cross-sections are taken from
Steidel 1993 and adapted to $H_o=75$~km~s$^{-1}$Mpc$^{-1}$. The
$z\approx 3$ galaxy is centered in a 5$''$ diameter circle that
subtends 37.5 kpc ($\Omega=0.2$).  The Holmberg diameter of
NGC628 is $\sim$36~kpc at a distance of 10~Mpc; the outermost contour
is $1.3{\times}10^{19}$cm$^{-2}$ and over half of the absorbing 
cross section is above $10^{20}$cm$^{-2}$.
}
\label{crosssection.fig}
\end{figure}

An interesting comparison can be made between the sizes of the
high-$z$ star forming galaxies (Giavalisco et al 1996a) and the
interception cross-sections for uv absorption by different ions
(cf. Steidel 1993).  The Lyman-break color-selection technique
for identifying the star forming galaxies produces candidates 
with a density on the sky of ${\sim}1$~arcmin$^{-2}$ for objects
with redshifts predominantly in the range $2.6\leq z\leq 3.4$
(Steidel et al 1998).
The comoving density of $L_*$ galaxy ``sites,'' computed for this redshift
range, amounts to ${\sim}3$~arcmin$^{-2}$ (for a cosmological model
with $\Omega_o=0.2$). Fig.~\ref{crosssection.fig} shows 
the  cross section for absorption lines  that every
$L_*$ galaxy site would necessarily present, if ordinary galaxies
are to explain the observed incidence of absorption lines.  
Thus, absorption line statistics indicate $\sim$3 times
the absorption cross sections shown in Fig.~\ref{crosssection.fig}
for every Lyman-break galaxy.
In fact, a few nearby  ${\sim}L_*$ galaxies observed in 21cm emission
at $z\approx 0$ present comparable cross sections.
The area for NGC628 with 
column densities above 10$^{20}$cm$^{-2}$ shown in the figure 
are comparable to the
DLa cross section, while the more extended emission at levels above $10^{19}$
compare with the required cross section for the MgII and Lyman-limit
selected systems. Tests with the UV image of NGC628 (Chen et al 1992), 
after simulation of redshifting and and cosmological dimming
(Giavalisco et al 1996b), lead to
the conclusion that even such a luminous galaxy would not
be detected in HST images beyond $z\sim 1$.
The implications for high-$z$ is that there is substantial gaseous content,
possibly accompanied by stellar emission, that can well go unnoticed
in deep optical images. The nature of these invisible absorbers remains
a puzzle.  Are the statistical cross sections built of small clouds that 
coalesce steadily since $z\approx 5$ to form the large galaxies at $z=0$
(Khersonsky \& Turnshek 1996), or are there already large bound structures, 
comparable to the cross sections of Fig.~\ref{crosssection.fig}
at high redshift?

\subsection{The Past: Redshifted 21cm Lines}

The pioneering observations for redshifted 21cm absorption-lines were
done at Green Bank.  These studies included ``blind'' spectral
surveys, where relatively narrow spectrometer bands were 
stepped through frequency to eventually cover large spectral
ranges and leading to the discovery of the first line, in 3C286,
by Brown \& Roberts (1973) and the line in 3C196 (Brown \& Mitchell 1983).
Selection on metal-line redshifts, such as MgII/MgI absorption, 
also led to success here in Green Bank (Roberts et al 1976;
Brown \& Spencer 1979).
The success rate of 21cm line detection at MgII selected redshifts
led to a statistical argument about the nature of the 21cm lines
at $<z>\sim 0.5$, namely,
that at a success rate of about one in ten, the 21cm line cross section
is roughly consistent with the area presented by the cool disks of
spiral galaxies.  In this picture, the MgII cross section is inflated
by the inclusion of lower column density (but still predominantly neutral)
gas in an extended halo. It can be seen in Fig.~\ref{crosssection.fig},
that the ``approximately one in ten'' statistic also holds 
for the DLa to MgII ratio.
 
A more reliable predictor of the detectability of the 21cm line
is the presence of strong DLa lines, since the necessary
high column density of HI is then clearly present (Wolfe and Davis 1979,
Wolfe, Briggs \& Jauncey 1981, Wolfe et al 1985, de Bruyn, O'Dea, 
\& Baum 1996), although it
also has become clear that high mean spin temperatures probably play
a role in lowering the observed 21cm optical depths (Taramopoulos et al 1995,
Carilli et al 1996, Taramopoulos et al 1996, Briggs, Brinks \& Wolfe 1997,
Kanekar \& Chengalur 1997) relative to the values measured for narrow
lines on lines of sight through the Milky Way and LMC (Dickey et al 1994).

New selection techniques have been successful, by, for example, 
observing redshifts of known
intervening gravitational lensing galaxies (Carilli, Rupen, \& Yanny
1993) and more recently by identifying highly reddened objects with
possible or likely gravitational lenses (Carilli et al 1998). 
For the highly reddened objects, it is not entirely clear in all
cases whether the reddening occurs in an intervening object or in
the background quasar host itself. In general, the full identification of 
complete samples of radio sources argues against intervening extinction
causing us to be missing vast populations of background quasars
(Shaver et al 1996).

At low redshifts, several 21cm absorption systems have been detected
in the outskirts of nearby galaxies, sometimes selected by
``proximity'' of the galaxy and quasar in the sky
 and sometimes through the presence of metal line absorption, 
such as Ca~H,K, against the quasar
 (Haschick \& Burke 1975, Carilli \& van Gorkom 1992). This work
is complemented by strenuous observational programs at optical and
uv wavelengths to measure absorption in extended disks, ionized halos,
and LSB dwarf companions of nearby galaxies against background objects
(Womble 1993, Bowen et al 1995).

Although the 21cm absorption line systems have been relatively few in
number  in comparison to the number of optical and uv lines, 
they have attracted a vigorous follow up effort, both
in ground based spectroscopy and imaging, as well as with HST
(Meyer \& York 1992, Lanzetta \& Bowen 1992,
Cohen et al 1994, Cohen et al 1996,  Burbidge et al 1996,
Le Brun et al 1997); since
Lyman-$\alpha$ is not shifted to wavelength accessible from the
ground until $z>1.7$, the low-$z$ 21cm absorbers were the most
accessible ``nearby'' examples of the DLa class of absorber.  More
recently, ultraviolet spectroscopic surveys of bright quasars
with HST are producing low $z$ DLa systems (Lanzetta et al 1997,
Turnshek 1997), which in turn make
interesting source lists for 21cm follow up. 

A number of VLBI experiments have been attempted in the
redshifted line with the goal of measuring transverse
structure and kinematics in the intervening absorbers
(Wolfe et al 1976, Johnston et al 1979, Brown et al 1988, Briggs
et al 1989).
Since these experiments required receiving systems for
the specific sky frequencies of the redshifted line, there have seldom
been more than two stations participating in each experiment, and
typically, one of the stations was
Arecibo, which has a limited tracking range, with the net result that
only very sparse uv coverage was obtained.  In principle, even this
limited coverage can be of use, since it does serve to resolve away
extended structure and isolate the
absorption against the core component, thus selecting the same
line of sight as probed by the optical/UV spectroscopy. In this way,
a fairer estimate of spin temperature can be obtained by comparing
the Lyman-$\alpha$ and 21cm line optical depths.

\section{The Present}

Progress in  observing redshifted lines has been limited in the past
by the frequency coverage available at the large fully-steerable 
telescopes and by radio interference. Recent installation
of wide-band receiving systems at the Westerbork Synthesis Radio
Telescope (WSRT) has led to a surge in the discovery of redshifted
21cm and OH lines,
as well as providing a fresh look at some old favorites.  Interferometers
are relatively invulnerable to moderate levels of interference, and 
the new observations are showing that the new absorption systems
are there, awaiting discovery. 

New classes of redshifted 21cm absorber have been turned up during the
last two years. The study of compact symmetric radio sources (CSOs) by
Vermeulen (this workshop) has shown that these sources as a rule 
do have associated 21cm absorption, apparently closely tied to the
nuclear region of the source. De Vries, O'Dea and colleagues are
finding  absorption associated with gigaHertz peaked radio
sources.
Lane (poster this workshop, Lane et al 1998), 
using MgII absorption to select redshifts
for observation, is finding new 21cm lines in roughly the
expected proportion.

\subsection{PKS~1229-021 revisited}

\begin{figure}
\psfig{figure=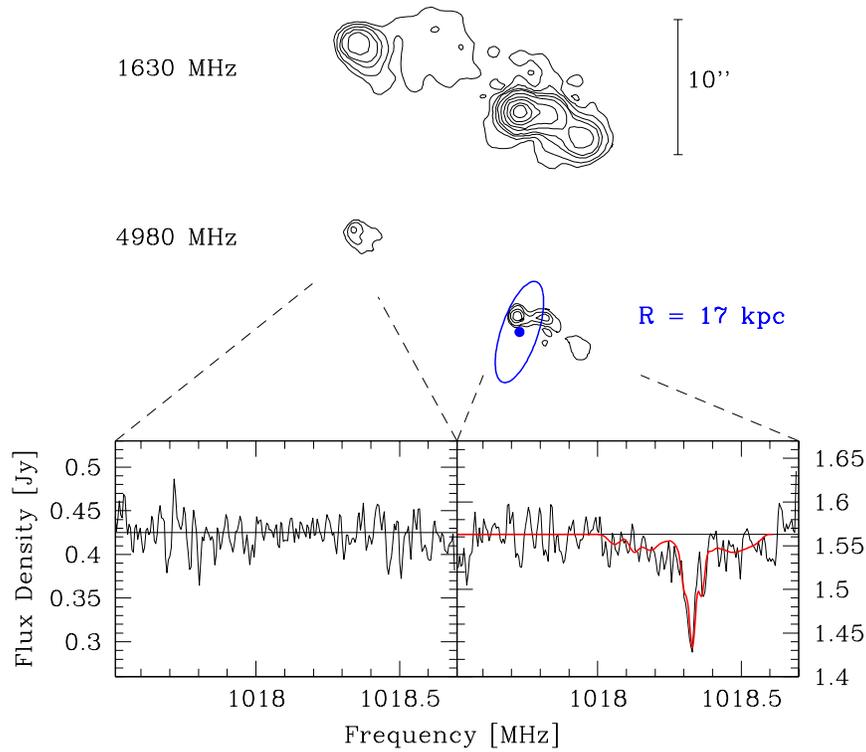,width=11cm,angle=0}
\caption{Background continuum structure of the $z_{em} = 1.045$
quasar PKS 1229-021 (Kronberg et al 1992) and the
intervening $z_{abs}=0.395$
21cm line absorption spectrum (Briggs, Lane and de Bruyn 
1998). {\it Top map}: 1630 MHz. {\it Lower map}: 4980 MHz. 
Spectra against the two
components can be isolated as indicated in the {\it lower panels}.
An oval is drawn over the SW component to represent the
kinematic model described in the text and Fig.~\ref{hst.fig}.
The simple model produces the smooth curve for the narrow and
broad absorption components in the right panel.}
\label{rad.fig}
\end{figure}

The quasar PKS 1229-02 presents
an interesting case study of the intervening galaxy class of 21cm line
absorber (Brown \& Spencer 1979). 
As one of the earliest identified absorption systems 
(Kinman \& Burbidge 1967) through its strong MgII doublet
at $z_{abs}=0.395$, it has received
much attention (Peterson \& Strittmatter 1978,
Brown \& Spencer 1979, Briggs \& Wolfe 1983), 
becoming a prototype of the DL$\alpha$ systems with
asymmetric metal lines (Briggs et al 1985, Prochaska \& Wolfe 1997) 
and the focus of
a detailed VLA polarization study by Kronberg, Perry \& Zukowski
(1992), since the radio continuum extent of ${\sim}15''$ in
the background quasar permits mapping the
Faraday rotation  along a cut through the disk of the
intervening absorber.  

The radio source is marginally resolved by the Westerbork synthesized
beam at 1018 Mhz, allowing the data to be decomposed to obtain
spectra against the two main components as shown in Fig.~\ref{rad.fig}.
The absorption appears to be concentrated on the SW structure, and
no absorption is detected against the NE extended lobe. In addition
to the narrow 21cm absorption that has been known for some time
(Brown \& Spencer 1979), the new WSRT data show broad,
low level absorption in 
the 21cm profile at frequencies both above and below the narrow line. 

\begin{figure}
\psfig{figure=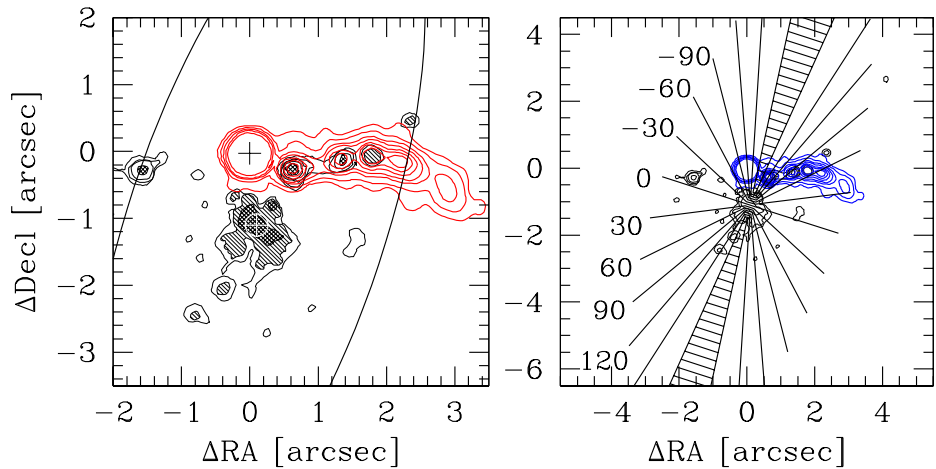,width=13cm,angle=0}
\caption{{\it Left}: HST image (Le Brun et al 1997). 
Contours with shading represent
the optical emission after subtraction of the QSO point-spread function, 
which is centered under the cross. Overlaid gray 
contours show the 8.2 GHz radio
continuum of the background QSO from P. Kronberg. The oval indicates the
size of a large disk galaxy as drawn in Fig.~\ref{rad.fig}.
{\it Right}: Differentially rotating
disk model for the distributed absorption, drawn
on a larger scale than in the left panel. Velocity
contours are drawn in 30 km~s$^{-1}$ intervals, with shading
used to highlight the $\pm$200 km~s$^{-1}$ ranges.}\label{hst.fig}
\end{figure}

Ground-based  (Steidel et al 1994) and HST imaging (Le Brun
et al 1997) have identified the absorber with a galaxy of optical
luminosity $L\ge 0.25 L_*$. 
HST spectroscopy has measured
the HI column density in the DLa line,
which permits estimation of the spin temperature (see also
Lane this workshop).
Comparison of the angular size of the radio structure with the angular
size subtended by a moderately bright disk galaxy at this redshift
(see Figs.~\ref{hst.fig} and \ref{rad.fig})
shows that the quasar nucleus must be close to the disk center and
that the radio jet extends on galactic scales behind the intervening
disk galaxy. 
  
Tests of a range of differentially rotating disk models describing gas-rich
disk galaxies show that an inclined disk with a flat rotation curve and
a gradually declining optical depth with radius mimics very well both the
narrow feature and broad wings in the absorption profile. The optical
depth against the quasar core is $\tau\approx 0.25$, and parameters
for inclination and position angle of the major axis are surprisingly
(see Fig.~\ref{hst.fig}) well constrained for such a model. 
The natural next step is
to map the absorption with a moderate resolution interferometer system
operating at 1018 MHz.

\section{Future}

The broadband spectrometers coming on line over the next year
will enable  radio spectral surveys for absorption against large 
samples of high $z$  sources. It is clear that the low
redshift regime of Fig.~\ref{fnct_z.fig} will benefit from
radio surveys to find DLa systems, free from bias by dust and
Lyman limits, and without need to resort to HST. Furthermore,
selection on
21cm absorption permits use of optically dim, high-$z$
radio galaxies as the background sources, with the advantage
that optical follow up to identify the intervening DLa absorber 
and study its environment will be simpler, since there are less
stringent requirements on precise subtraction of
the quasar point-spread function. Selection on 21cm can detect
heavy absorption against extended radio components in quasars for
which the optical nucleus is uncovered and shows no DLa line; in this sense,
the largest high-$z$ radio sources provide multiple lines of sight, increasing
the effiency of observing the high $N_{HI}$ end of the absorption-line
distribution.

\begin{figure}
\psfig{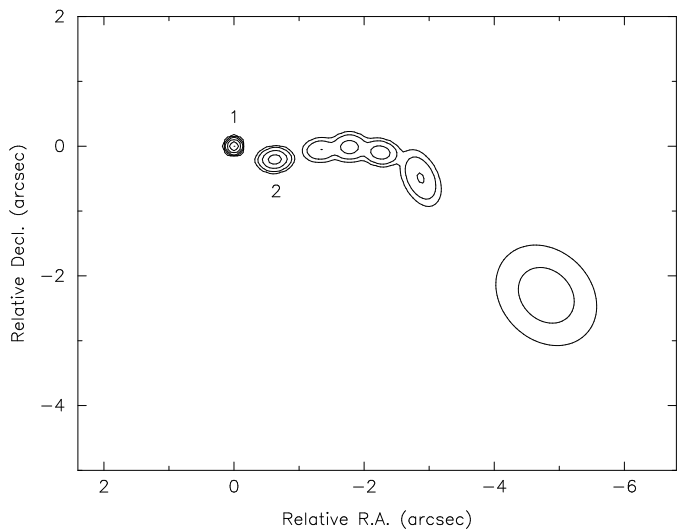}
\caption{PKS 1229-02 model for VLBI observations at 1018 MHz. The map was
constructed with 0.1$"$
resolution and contour levels at 0.5, 2, 10, 40, 200, 600 mJy per
beam. Integral fluxes of each of the model components along the jet are
 600, 250, 110, 115, 125, 300, 430 mJy from left to right.
}\label{contourmap.fig}
\end{figure}

\begin{figure}
\psfig{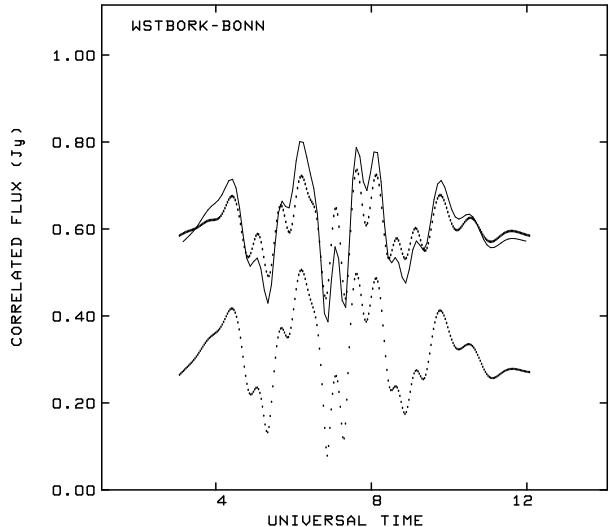}
\caption{PKS 1229-02 visibility as a function of time for a 
Westerbork-Effelsberg baseline at 1018 MHz. Visibility in the
continuum is indicated by the solid line. The {\it lower dotted}
line shows the effect of halving the flux density of component
1 in Fig.~\ref{contourmap.fig}.
The {\it upper dotted}
line shows the effect of halving the flux density of component
2.}\label{visib.fig}
\end{figure}

High spatial resolution observations will define the extent and fine-scale
kinematics in intervening absorbers as well as measuring the local 21cm
optical depth against the background nucleus for comparison with
DLa lines and refinement of spin temperatures. To this end, a number
of EVN telescopes are being equipped with UHF receivers, and test
observations are showing that this kind of work can be done despite
levels of rfi that would be intolerable in single-dish spectroscopy.

As an example of how very, limited u-v coverage can provide
kinematic information, Fig.~\ref{contourmap.fig} shows a simple
model for PKS1229-021, made by adjusting the flux densities of the
knots seen in Fig.~\ref{hst.fig} to 1018 MHz, while keeping the
positions and shapes fixed.  Fig.~\ref{visib.fig}
illustrates the response of a single Westerbork-Bonn baseline to the
continuum model, along with models where selected components
attenuated, to simulate localized, frequency dependent absorption.
The complex visibility ${\bf V}(u,v)$ is
modeled as the sum of the visibilities ${\bf V}_i$ of the individual
components with a phase factor to account for each component's
position ($x_i,y_i$) in the sky:
$${\bf V}_c(u,v) 
= \sum {\bf V}_i(u,v)e^{i2\pi(ux_i+vy_i)} $$
Each component appears in the sum with
a distinctive signature (pseudo-periodicity) defined by the source model. 
Absorption only changes the strength of a component in the sum 
(${\bf V}(u,v,\nu) 
= \sum f_i(\nu){\bf V}_i(u,v)\exp[{i2\pi(ux_i+vy_i)}]$),
so by fitting the visibilities to find 
the relative strengths $f_i(\nu)$ at every frequency,
a spectrum for each component can be constructed.

A variety of follow-up observations are possible for the growing
sample of 21cm lines.
Faraday rotation (cf Kronberg et al 1992) and Zeeman splitting in
very narrow and deep lines will soon be specifying magnetic fields
in the absorbers.
The time has come to re-examine the idea of Davis \& May (1978) that
precise measurements of narrow radio absorption lines (in HI or molecules)
could determine  fundamental constants ($\Omega_o$ and
$\Lambda$) by observing the deacceleration of the Universe as a function
of $z$ in observations spaced over a few decades.


\end{document}